\documentclass[aps,preprint,superscriptaddress]{revtex4}

\usepackage{amsmath} 
\usepackage{amssymb} 
 \usepackage{amsfonts}
\usepackage[dvips]{graphicx} 
\usepackage[]{epsfig} 
\usepackage[normalem]{ulem}

\begin{document}

\bibliographystyle{apsrev} 

\title{Dewetting-controlled binding of ligands to hydrophobic pockets} 

\author{P. Setny} 
\affiliation{Department of Chemistry and Biochemistry,  UC San Diego, La Jolla, CA 92093, USA}
\affiliation{Interdisciplinary Center for Mathematical and Computational Modeling, University of Warsaw, Warsaw 02-089, Poland}
\author{Z. Wang}
\affiliation{Department of Chemistry and Biochemistry,  UC San Diego, La Jolla, CA 92093, USA}
\affiliation{Department of Mathematics, UC San Diego, La Jolla, CA 92093, USA}
\author{L.-T. Cheng}
\affiliation{Department of Mathematics, UC San Diego, La Jolla, CA 92093, USA}
\author{B. Li}
\affiliation{Department of Mathematics, UC San Diego, La Jolla, CA 92093, USA}
\affiliation{NSF Center for Theoretical Biological Physics (CTBP), 
UC San Diego, La Jolla, CA 92093, USA}
�\author{J.~A. McCammon}
\affiliation{Department of Chemistry and Biochemistry,  UC San Diego, La Jolla, CA 92093, USA}
\affiliation{NSF Center for Theoretical Biological Physics (CTBP), 
UC San Diego, La Jolla, CA 92093, USA}
\affiliation{Department of Pharmacology and  HHMI, UC San Diego, La Jolla, CA 92093, USA} \author{J. Dzubiella}
\affiliation{Physics Department, Technical University Munich, 85748 Garching, Germany}
\thanks{Electronic address: jdzubiel@ph.tum.de}

%\maketitle

%\begin{article}

\begin{abstract}
We report on a combined atomistic molecular dynamics simulation and implicit solvent analysis of a generic hydrophobic pocket-ligand (host-guest) system. The approaching ligand induces complex wetting/dewetting transitions in the weakly solvated pocket. The transitions lead to bimodal solvent fluctuations which  govern magnitude  and range of the pocket-ligand attraction.  A recently developed implicit water model, based on the minimization of a geometric functional, captures the sensitive aqueous interface response to the concave-convex pocket-ligand configuration semi-quantitatively. 
  
\end{abstract}

\pacs{61.20.p, 68.03.g, 82.60.Lf, 87.15.v}

\maketitle 

The {\it water-mediated} interaction between a ligand and a hydrophobic binding pocket  plays a key role in biomolecular assembly processes, such as protein-ligand recognition~\cite{uchida:jbc:97,  Carey:2000p1788, Levy:2006p1071, young,Ahmad:2008p1785,qvist},  the binding of the human immunodeficiency virus (HIV) ~\cite{Braaten:1997p1775} or the dengue virus~\cite{Modis:2003p1786} to human cells,  the inhibition of influenza virus infectivity~\cite{modis},  or in synthetic host-guest systems~\cite{Gibb:2002p1790}.   Experiments and explicit-water molecular dynamics (MD) simulations suggest that the concave nature of the host geometry imposes a strong hydrophobic constraint and can lead to very weakly hydrated pockets~\cite{uchida:jbc:97,  Carey:2000p1788, Levy:2006p1071, young,Ahmad:2008p1785, qvist, Chandler:2005p1796, Berne:2009p1776},  prone to nanoscale {\it capillary evaporation} triggered by the approaching ligand~\cite{Ahmad:2008p1785,young,Setny:2007p470}.  This so called 'dewetting' transition has been also observed in other (protein) geometries~\cite{Chandler:2005p1796,Berne:2009p1776}.  It has been speculated that dewetting may lead to a fast host-guest recognition accelerating the hydrophobic collapse and binding rates of the ligand into its pocket~\cite{uchida:jbc:97, young,Ahmad:2008p1785}.  A deeper physical understanding of these sensitive hydration effects in hydrophobic recognition  is still elusive.

On the coarse-grained modeling side,  the thermodynamics of molecular recognition is typically approached by surface area (SA) models~\cite{roux:biochem}. A major flaw of these {\it implicit solvent} models is that the aqueous interface around the macromolecules is predefined (typically by rolling a probe sphere over the van der Waals surface) and is therefore a rigid object that cannot adjust  to local energetic potentials and changes in spatial molecular arrangements. In particular, the dewetting transition, which is highly sensitive to local dispersion, electrostatics, and geometry~\cite{Chandler:2005p1796,Dzubiella:2006p469,Berne:2009p1776}, can, {\it per definitionem}, not be captured by SA type of models. Their qualitative deficiency to describe the hydrophobic pocket-ligand interaction in proteins~\cite{Michel:2006p1792},  pocket models~\cite{Setny:2007p470},  or dewetting in protein folding~\cite{Rhee:2004p1793} is therefore not surprising. 

In this letter, we combine explicit-water MD simulations and the variational implicit solvent model (VISM)~\cite{Dzubiella:2006p469} applied to a generic pocket-ligand model. The simulations show that the approaching ligand first slightly stabilizes  the wet state in the weakly hydrated pocket, whereas, upon further approach, bimodal fluctuations in the water occupancy of the  pocket are induced, followed by a complete pocket dewetting. The onset of fluctuations defines the critical range of pocket-ligand attraction. The VISM calculation, based on the minimization of a geometric functional, reproduces the bimodal hydration and explains it by the existence of distinct metastable states which correspond to {\it topologically} different water interfaces. As opposed to previous (SA type of) implicit  models, VISM captures the range  of the pocket-ligand interaction semi-quantitatively.  Strikingly, the observed nanoscale phenomena can be thus explained by geometric capillary effects, well-known on macroscales~\cite{degennes}. Explicit inclusion of dispersion  interactions and curvature corrections, however, seem to be essential for an accurate description on nanoscales.  

Our generic pocket-ligand model consists of a hemispherical pocket embedded in a rectangular wall composed of  neutral Lennard-Jones (LJ) spheres interacting with $U_{\rm LJ}(r)=4\epsilon[({\sigma}/{r})^{12}-({\sigma}/{r})^6]$. The atoms are  aligned in a hexagonal closed packed arrangement with a lattice constant of 1.25~\AA.   The LJ parameters are chosen to model a paraffin-like material and are $\epsilon_{p}=0.03933$~kJ/mol and $\sigma_{p}=4.1$~\AA~~\cite{Setny:2007p470,SI}. We consider two different pocket radii: $R=5$ and 8~\AA, which we refer to in the following as 'R5' and 'R8' systems. The ligand is taken as a methane (Me) represented by a neutral LJ sphere with parameters $\epsilon=0.294$~kJ/mol and $\sigma=3.730$~\AA. It is placed at a fixed distance $d$ from the flat part of the wall surface ($z=0$), along the pocket symmetry axis in $z$-direction, see Fig.~1~(a) for an illustration. The explicit-water MD simulations are carried out with the CHARMM  package ~\cite{charmm} employing the TIP4P water model in the $NVT$ ensemble, two anti-symmetric walls with thickness of 7.5~\AA~in a surface-to-surface distance of $30$~\AA~in a rectangular box with lengths $L_{x}=L_{y}\simeq 34$~\AA~ and $L_{z}=100$~\AA, and 3D particle mesh Ewald summation. In equilibrating simulations, the volume $V$ of the system was varied until the density in the center of the slab matched the bulk density of TIP4P water at a pressure of $P=1$~bar and temperature $T=298$~K. More technical details of the simulation setup can be found elsewhere~\cite{Setny:2007p470,SI}.  A MD simulation snapshot is shown in Fig.~1~(b). 

The VISM was introduced in detail previously~\cite{Dzubiella:2006p469} and applied to the solvation of nonpolar solutes~\cite{Cheng:2007p442}. Briefly, let us define a subregion $\cal V$ {\it void} of  solvent in total space $\cal W$, for which we assign a volume exclusion function  $v(\vec r)= 0 $ for $\vec r \in \cal V$ and $v(\vec r)= 1 $  else. The volume $V$ and interface area $S$ of $\cal V$ can then be expressed as functionals of $v(\vec r)$ via $ V[v]=\int_{\cal W}{\rm d}^3r \;[1-v(\vec r)]$ and $S[v]=\int_{\cal W}{\rm d}^3r \;|\nabla v(\vec r)|=\int_{\cal \partial W}{\rm d}S$,  and the solvent density is  $\rho(\vec r)=\rho_0 v(\vec r)$, where $\rho_0$ is the bulk value. The solvation free energy $G$ is defined as a {\it functional} of the geometry $v(\vec r)$ of the form~\cite{Dzubiella:2006p469} 
\begin{eqnarray}
G[v] &=& P V[v]+\int_{\cal \partial W}{\rm d}S\;\gamma_{lv}[1-2\delta H(\vec r)] \nonumber \\
&+& \rho_0\int_{\cal W}{\rm d}^3r\;v(\vec r) U(\vec r), 
\label{eq:grand}
\end{eqnarray}
where $\gamma_{lv}$ is the liquid-vapor interface tension, $\delta$ the coefficient for the  curvature correction of  $\gamma_{lv}$ in mean curvature $H(\vec r)$, and $U(\vec r)=\sum_{i}^{N_{s}} U^{i}_{\rm LJ}(\vec r-\vec r_{i})$  sums  over the LJ interactions of all $N_{s}$ solute atoms at $r_{i}$ (ligand+wall atoms) with the water.  The $\delta$-term in (1) has been used in scaled-particle-theory~\cite{spt} for convex solutes only, generalized capillary theory~\cite{boruvka}, and in the morphometric approach applied to the solvation of model proteins~\cite{mecke3}. The minimization $\delta G[v]/\delta v=0$ leads to the partial differential equation (PDE)~\cite{Dzubiella:2006p469} 
\begin{eqnarray}
P-2\gamma_{{\rm lv}}\left[H(\vec r)+\delta K(\vec
r)\right]-\rho_0 U(\vec r)=0
\label{diff}
\end{eqnarray}
which is a generalized Laplace equation of classical capillarity~\cite{degennes, boruvka} extrapolated to microscales by dispersion and the local Gaussian curvature $K(\vec r)$.  The PDE (2) is solved using the {\it level-set method} which relaxes the functional (1) by evolving a 2D interface in 3D  space and robustly describes topological changes, such as volume fusion or break-ups~\cite{OsherFedkiw02, Cheng:2007p442}.  The free parameters chosen to match the MD simulation are $P=1$~bar, $\gamma_{lv}=$ 59~mJ/m$^{2}$ for TIP4P water~\cite{vega},  and $\rho_{0}=0.033$~\AA$^{-3}$. The coefficient $\delta$ is typically estimated to be between 0.8 and 1~\AA~for  various water models around convex geometries~\cite{Chandler:2005p1796,sokolov}, while VISM was able to predict well the solvation free energies of simple solutes for  $\delta=1$~\AA~\cite{Cheng:2007p442} which we use in the following.

We consider ligand positions from $d=11$ \AA$ $ to the distance of nearest approach to the pocket bottom. The latter is defined as corresponding to a wall-ligand interaction energy of 1~$k_{B}T$ and is $d\simeq -1.8$ and -3.8 ~\AA~ for the R5 and R8 system, respectively.  We define the water occupancy $N_{w}$ of the pocket by the number of oxygens whose LJ centers are located at $z<0$.  Considering the probability distribution $P(N_{w})$, we obtain the free energy as a function of $N_{w}$ by $G(N_{w})=-k_{B}T\ln P(N_{w})+G'$. Without the ligand (effectively for $d\gtrsim9$~\AA), the MD simulation reveals that the R5 pocket is in a stable dry state with occupancy $N_{w} \simeq N_{\rm dry}=0$, despite the fact that a few water molecules fit in and consistent with experiments on an equally sized protein pocket~\cite{qvist}. The R8 system, however, is found to be weakly hydrated. The $G(N_{w})$ distribution shown  for $d=9$~\AA~in Fig.~2  reveals an almost barrierless transition between wet and dry states.  The metastable wet state comprises  $N_{w}\simeq 9=N_{\rm wet}$ water molecules which roughly corresponds to bulk density. 

The approaching ligand considerably changes the $G(N_{w})$ distribution in the R8 system. As plotted in Fig.~2, for $d=6.5$~\AA~  the free energy exhibits a minimum at the wet state which is slightly stabilized (by $\simeq 0.4\,\, k_{B}T$) over the dry state. The function $G(N_{w})$ develops, however, concave curvature for $N_{w}\simeq 0$ indicative of the onset of a thermodynamic instability.  Indeed, upon further approach of the ligand ($d=5.5$~\AA) a local minimum forms at the dry state. It becomes a stable, global minimum at the critical distance $d_{c}\simeq4.5$~\AA. The now metastable wet state completely vanishes for $d\lesssim 0$~\AA, where we find a free energy difference between the wet and dry state of $G(N_{\rm dry})-G(N_{\rm wet})\simeq 5k_{B}T$. By investigating the water density distribution (Fig.~2, right panel), we find that a possible reason for the weakly stabilized wet state at $d=6.5$~\AA~ may be the first methane hydration shell partly penetrating the pocket. (This perhaps surprising effect should not be assigned to a lack of hydrophobicity or even an hydrophilic nature of the methane but to subtle hydrogen bond arrangements within this  geometry.) The average occupancy profile $\langle N_{w}(d) \rangle$  thus exhibits a maximum  at $d=6.5$~\AA~(where $\langle N_{w} \rangle \simeq 6$) while it jumps down to $\simeq 0$ at $d\simeq d_{c}$~\cite{SI}.

In the VISM  where thermal interface fluctuations are not yet considered, we start the numerical relaxation of the functional (1) from ({\bf i}), one closed solvent boundary which is arbitrary and loosely envelopes both the pocketed wall and the ligand, or ({\bf ii}),  the (tight) van der Waals surface around the wall and the ligand giving rise to two separated surfaces.  In Fig.~3 we plot examples of the resulting VISM interfaces for both (i) and (ii), obtained for the ligand at $ d=4.5$~\AA.  For (i) the solution relaxes to a single interface that wraps both wall and the ligand together, thereby indicating a dry pocket state [Fig.~3~(a)], while for (ii) the solution relaxes to two separate surfaces, one of which closely follows the pocket contours indicating a wet state [Fig.~3~(b)]. The existence of two distinct results can be clearly attributed to the energy barrier between wet and dry states observed in the simulation (cf. Fig.~2).

By systematically investigating different initial configurations and ligand distances we find that the solution for R8 relaxes to at most three distinct interfaces: 1. a single enveloping surface around the dry pocket and ligand ({\bf 1s}), 2. two separated surfaces with a dry pocket ({\bf 2s-dry}), and 3. two separated surfaces with a wet pocket ({\bf 2s-wet}).  Selected examples for the interface at ligand distances $d=-2,2,4.5$ and 9~\AA~are shown in Fig.~3, where we plot the bisected VISM interface for a clearer view.  For the initial configuration (i) and for $d\lesssim 7$~\AA, the results converge to the 1s state while for larger separations a breakup into two interfaces (2s-dry) is observed [Fig.~3~(c)].  The stable 2s-dry state exists also for $5<d<7$~\AA, where it is reached from an initial configuration intermediate between (i) and (ii). For the initial configuration (ii) and for $d\gtrsim 0$, the results converge to two separated surfaces with a wet pocket (2s-wet) while for smaller separations there is only one enveloping surface (1s), see Fig.~3~(d). For R5 we just find two distinct solutions, 1s and 2s-dry, indicating a very stable dry pocket in agreement with the MD simulations and experiments~\cite{qvist}. These results demonstrate that VISM captures the dewetting transition, and the final interface is relaxed into (meta)stable states representing (local) free energy minima. This is in physical  agreement with the bimodal behavior observed  in the MD simulation and is further quantified in the following. 

The minimum VISM free energy (1) vs. $d$ is is plotted for R8 in Fig.~4: for $d<0$~all possible VISM solutions converge to 1s, featuring a dry pocket. For $0\lesssim d\lesssim d_{c}\simeq4.5$~\AA, the 'dry branch' 1s is favored over the second appearing branch corresponding to the 2s-wet interface (by $\simeq$ 8 $k_{B}T$ at $d=0$) in excellent agreement with the two-state behavior in the MD simulation.  For $d_{c}\lesssim d\lesssim 7$~\AA~ the 2s-dry state is favored over 2s-wet and 1s which is now highly metastable.  For $d\gtrsim 7$~\AA ~the 1s state disappears and 2s-dry is favored by roughly 2$k_{B}T$  over 2s-wet. The fact that a dry pocket is favored in VISM for large $d$ is in contrast to the MD simulation which supplied a very weakly hydrated pocket for $d\gtrsim 6.5$~\AA.  Changing the curvature parameter $\delta$  shows that this failure can be attributed to a too high energy penalty for concave interface curvature (a too large $\delta$ for $H<0$) which favors pocket dewetting.  It thus appears that the simple curvature correction applied breaks down and is  {\it not symmetric} with respect to concavity and convexity on these small scales. The symmetry may be broken by higher order correction terms in the the curvature expansion of the surface tension, if feasible~\cite{evans2}.

If thermal fluctuations were included in VISM, the various energy branches would be sampled in a Boltzmann-weighted fashion to yield the solvent-mediated potential of mean force (pmf)  between the ligand and the pocket.  At present, allowing the existence of multiple local minima for a given $d$ in the $G[v]$ functional that correspond to the ensemble $\{v\}_m$ of most probable solvent configurations, we obtain the ensemble average (EA) as $G= -k_{B}T \ln \sum_{\{v\}_m}e^{-G[v]/k_{B}T}+G''$, with $G''$ being an arbitrary constant. The resulting pmfs $G(d)$ for R8 and R5 are shown in Fig.~4 together with the MD simulation results. The curves are overall in good, almost quantitative agreement.  A detailed analysis of the individual energy contributions to (1)  reveals that the inclusion of the dispersion and curvature correction terms in VISM is crucial to capture the onset of the attraction at $d_{c}=4.5$~\AA~for R8, while SA type of calculations yield a too low $d_{c}\simeq 0$~\cite{Setny:2007p470}.  Furthermore,  the $\sim 1~k_{B}T$ energy barrier at $d\simeq 6$~\AA~ for the R5 system is nicely captured by VISM; it can be attributed  to the unfavorable curvature correction term arising from the development of a concave solvent boundary penetrating the pocket, as well as the wall-water dispersion term, whose repulsive contribution stems from displacement of water close to the small R5 pocket.  An EA performed to estimate the average occupancy $\langle N_{w}(d) \rangle$  for R8 yields qualitative agreement with the MD, i.e., a step at $d\simeq d_{c}$ from zero to nonzero occupancy and a  maximum at $d\simeq 6.0$~\cite{SI}.  The step height in VISM ($\Delta N_{w}\simeq 2$), however, is much smaller as in the simulation ($\Delta N_{w}\simeq 6$) which is probably due to the continuum approximations of VISM and the neglect of fluctuations in performing the EA.

In summary, the geometry-based VISM is the first implicit solvent model that captures the multi-state hydration observed  in simulations and experiments ~\cite{Berne:2009p1776} and highlights the significance of interfacial fluctuations~\cite{mittal} in hydrophobic confinement where the free energy can be polymodal.  Pocket dewetting may be regarded as the rate-limiting step for protein-ligand binding as found in folding~\cite{Chandler:2005p1796}. The existence and height of activation barriers and the range of attraction, however, can strongly depend on pocket size and geometry in terms of local interface curvature. Our capillary approach, where dispersion  and curvature effects play explicit roles, may represent a valuable step towards proper interpretation and modeling of experimental binding rates.~\cite{uchida:jbc:97}.

BL is supported by the National Science Foundation (NSF), Department of Energy, 
and the CTBP. JAM is supported in part by NIH, NSF, HHMI, NBCR, and
CTBP.  JD thanks the  Deutsche Forschungsgemeinschaft (DFG) for support 
within the Emmy-Noether-Program.

\newpage 

\begin{figure}[h]
\includegraphics[width=12cm,angle=0]{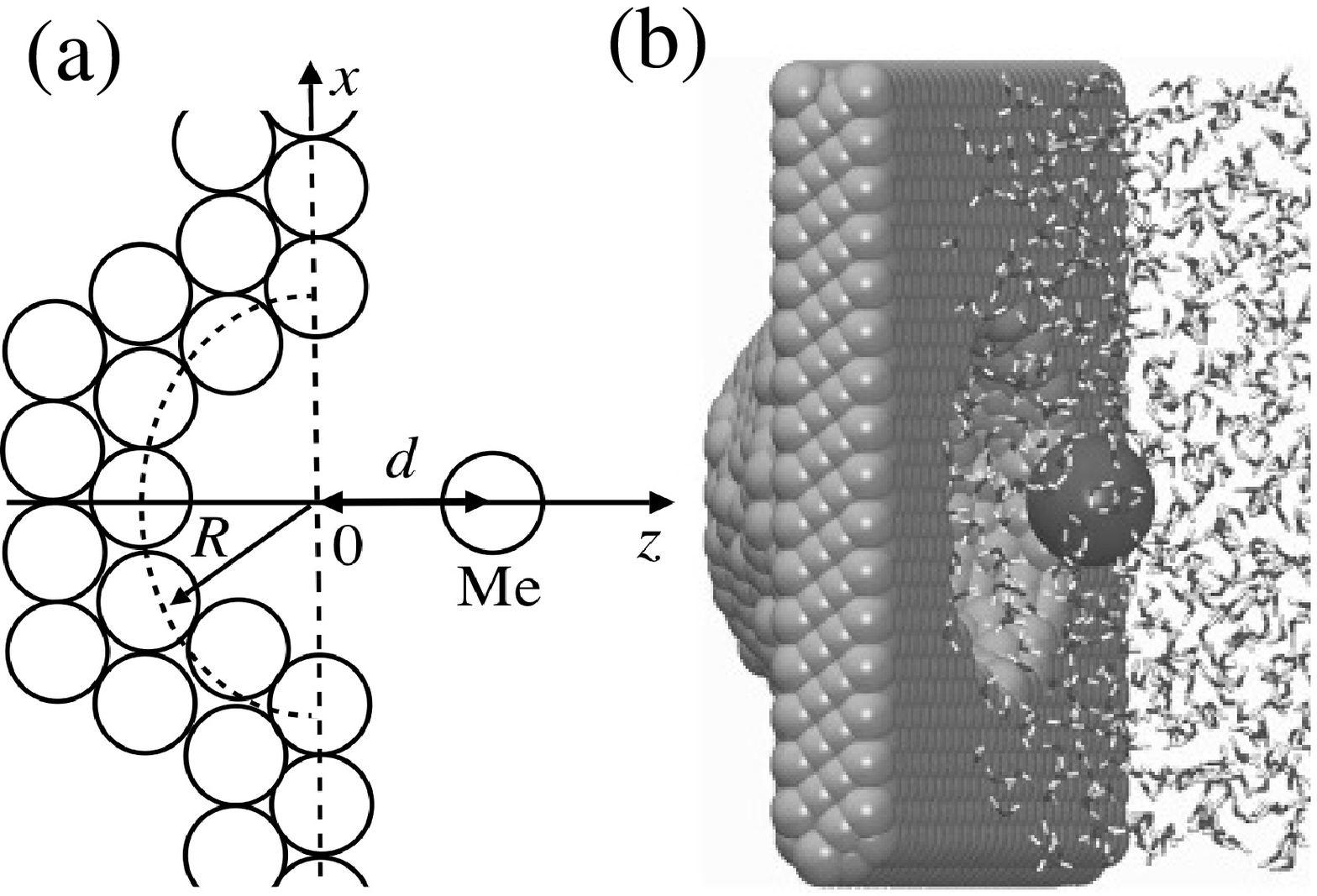}
\caption{ (a) Sketch of the generic model. The pocket has a radius $R$.  
The methane (Me) ligand  is fixed at a distance $d$ from the wall surface. (b) MD simulation snapshot illustrating the wall/ligand/water system.}
\label{fig1}
\end{figure}

\begin{figure}[h]
\includegraphics[width=12cm,angle=0]{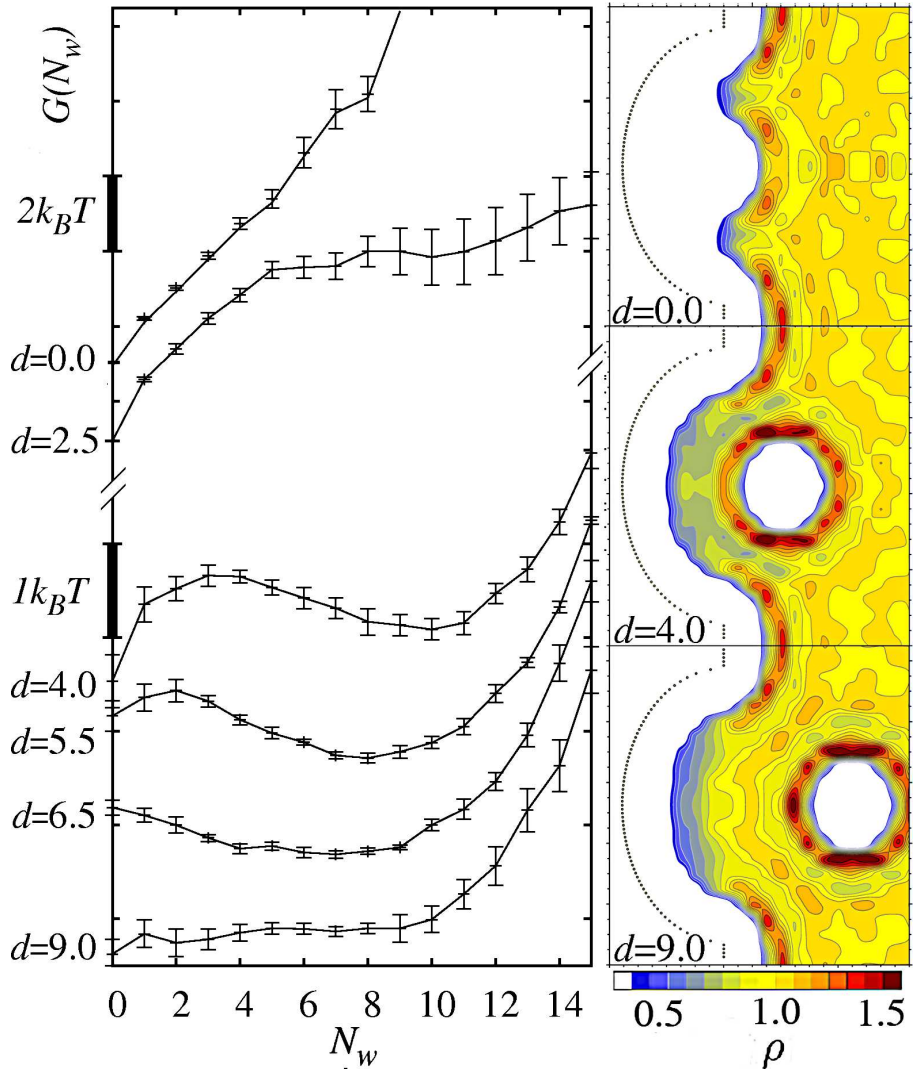}
\caption{MD simulation results for the free energy $G(N_{w})$ vs. pocket water occupancy $N_{w}$ in pocket R8 for ligand distances $d$=0.0, 2.5, 4.0, 5.5, 6.5, and 9~\AA. The curves are shifted vertically and we use two scales (1 and 2 $k_{B}T$) for a better illustration.  The right panel exemplifies the water density ($\rho$) distribution around pocket and ligand for selected~$d$.}
\label{fig2}
\end{figure}

\begin{figure}[h]
\includegraphics[width=12cm,angle=0]{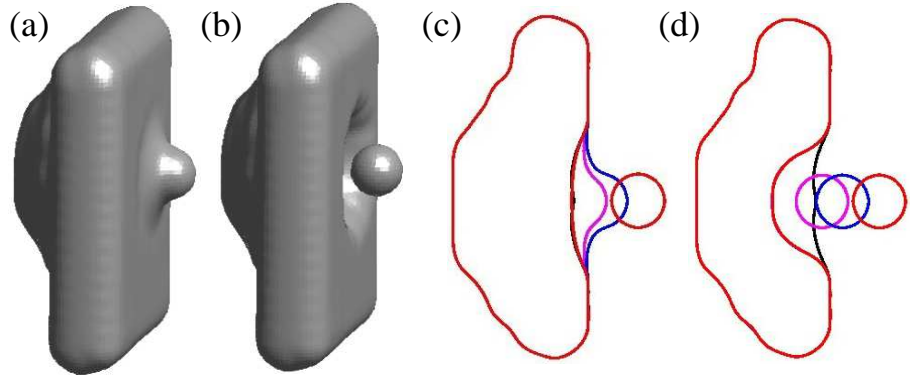}
\caption{ VISM solution of the aqueous interface for the R8 system. a) and b), full three-dimensional interface for the ligand at $d=4.5$ for one surface (i) and two separated surfaces (ii) as initial boundary inputs, respectively. c) and d), the bisected interface for initially one surface (i) and two surfaces (ii), respectively,  for distances $d=$ -2, 2, 4.5, and 9~\AA~(black, magenta, blue, and red).}
\label{fig3}
\end{figure}

\begin{figure}[h]
\includegraphics[width=12cm,angle=0]{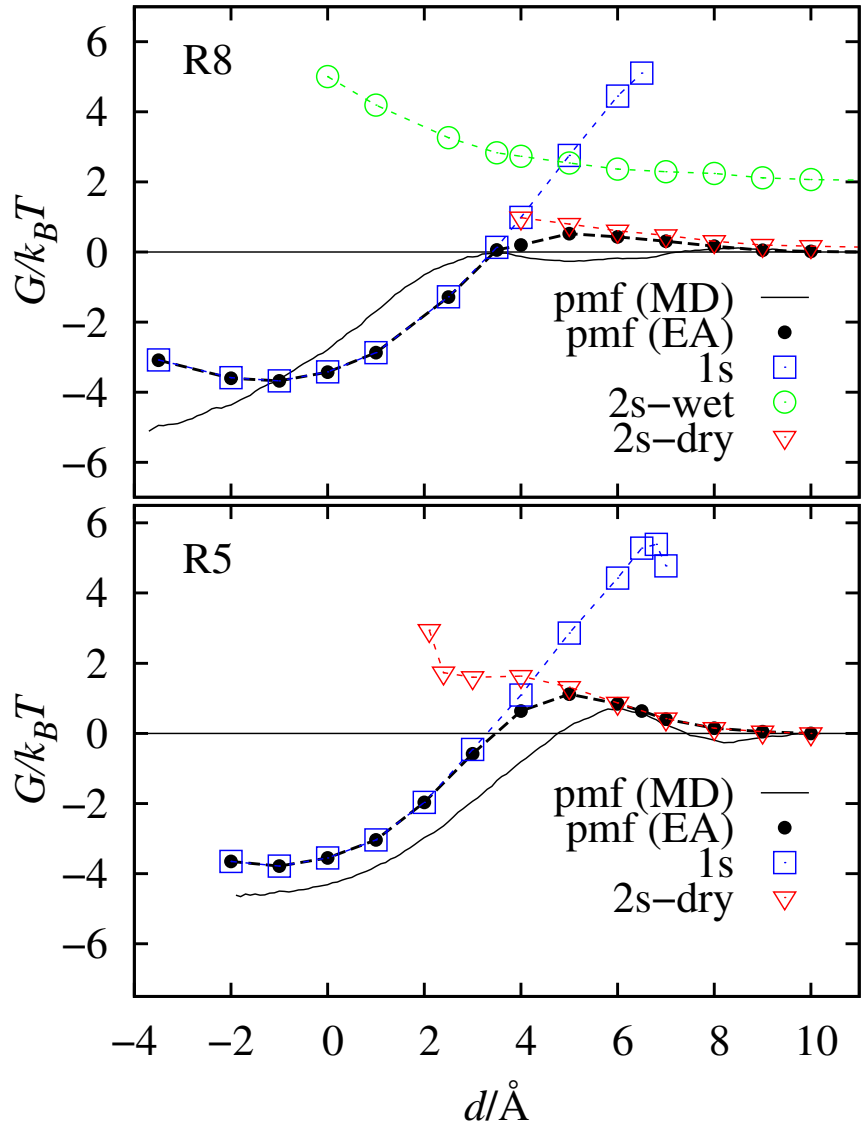}
\caption{VISM free energies for the 1s (squares),  2s-wet (circles), and 2s-dry (triangles) branches for R8 (top) and R5 (bottom) and the solvent-mediated pmf between the pocket and ligand from MD simulations (solid lines) and the ensemble average (EA) over the VISM branches (filled circles). }
\label{fig4}
\end{figure}

 \end{document}

\bibliographystyle{apsrev} 

\title{Supporting Information} 

\author{P. Setny} 
\affiliation{Department of Chemistry and Biochemistry,  UC San Diego, La Jolla, CA 92093, USA}
\affiliation{Interdisciplinary Center for Mathematical and Computational Modeling, University of Warsaw, Warsaw 02-089, Poland}
\author{Z. Wang}
\affiliation{Department of Chemistry and Biochemistry,  UC San Diego, La Jolla, CA 92093, USA}
\affiliation{Department of Mathematics, UC San Diego, La Jolla, CA 92093, USA}
\author{L.-T. Cheng}
\affiliation{Department of Mathematics, UC San Diego, La Jolla, CA 92093, USA}
\author{B. Li}
\affiliation{Department of Mathematics, UC San Diego, La Jolla, CA 92093, USA}
\affiliation{NSF Center for Theoretical Biological Physics (CTBP), 
UC San Diego, La Jolla, CA 92093, USA}
�\author{J.~A. McCammon}
\affiliation{Department of Chemistry and Biochemistry,  UC San Diego, La Jolla, CA 92093, USA}
\affiliation{NSF Center for Theoretical Biological Physics (CTBP), 
UC San Diego, La Jolla, CA 92093, USA}
\affiliation{Department of Pharmacology and HHMI,  UC San Diego, La Jolla, CA 92093, USA}
\author{J. Dzubiella}
\affiliation{Physics Department, Technical University Munich, 85748 Garching, Germany}
\thanks{Electronic address: jdzubiel@ph.tum.de}

\maketitle

%\begin{article}

\subsection{The wall-water interaction}
\begin{figure}[h!]
\includegraphics[width=8cm,angle=0]{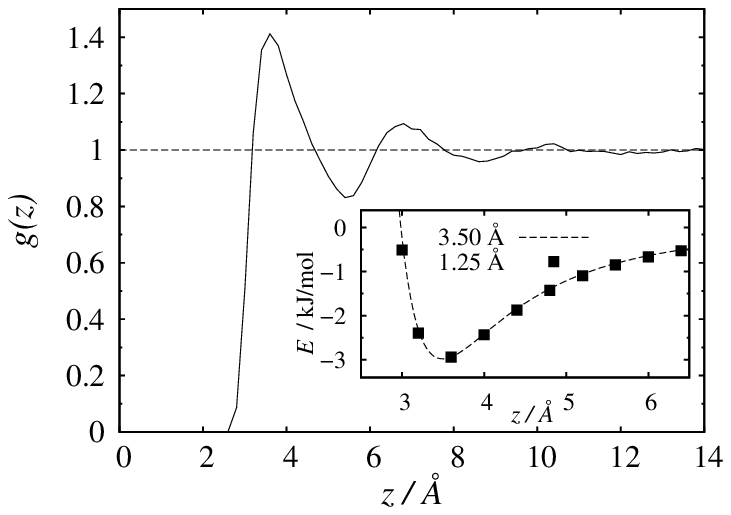}
\caption{Water oxygen density vs. the distance $z$ from the flat wall surface; $g(z)=1.0$ corresponds to a water number density of 0.0334 \AA$^{-3}$. Inset: wall-water interaction energy for the original 3.5 \AA$ $ grid lattice (dashed line), and the 1.25 \AA$ $ grid lattice with adjusted LJ potential parameters (squares). }
\label{fig1}
\end{figure}

In order to construct hydrophobic walls we considered a paraffine-like material of 0.8 g/cm$^3$ density composed of CH$_2$ units. Assuming a hexagonally closed-packed arrangement, the given density requires a lattice constant of 3.5 \AA~ which is too coarse to produce a relatively smooth hemispherical pocket. Thus, we reduced the lattice constant to 1.25~\AA~while at the same time adjusting the Lennard-Jones potential parameters of the wall pseudo-atoms to reproduce the original paraffine wall - water interaction energy (see inset to Fig.~\ref{fig1}) that was obtained with the united atom OPLS parameters for CH$_2$ units~[1]. The wall-water interaction energy was calculated by simply averaging the interaction energy over planes at constant $z$. 

The height of the first peak in the wall-water density profile from our MD simulations (Fig.~\ref{fig1}) is within the range (1.3 to 1.6) observed in all atom MD simulations of hydrocarbon-water interfaces~\cite{lee_1994,jensen_2004,pal_2005a}, suggesting that the walls indeed closely resemble a paraffine-like material.

\subsection{Average water occupancy of the pocket}
\begin{figure}[h!]
\includegraphics[width=8cm,angle=0]{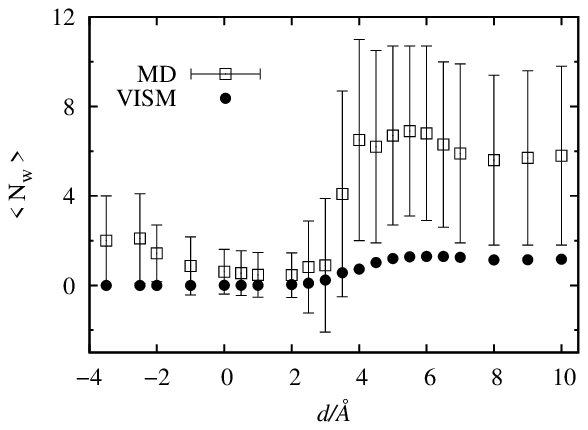}
\caption{Average pocket occupancy $\langle N_{w}\rangle$ from MD simulation and VISM ensemble average.}
\label{fig2}
\end{figure}

Based on the VISM results we estimated an average pocket occupancy $\langle N_w \rangle$ for different ligand positions by the ensemble average
% EA for $\langle N_{w}\rangle$: 
\begin{equation}\label{ea}
\langle N_{w}\rangle= \sum_{\{v\}_m}N[v]e^{-G[v]/k_{B}T}/\sum_{\{v\}_m}e^{-G[v]/k_{B}T},
\end{equation}
where $N[v]=0$ for dry-type solutions and $N[v]=N_{wet}=9$ for wet-type solutions.
A comparison with MD results (Fig.~\ref{fig2}) shows the correct qualitative behavior with a transient increase in pocket wetting at $d\simeq6$~\AA, followed by %a sharp decrease of average occupancy due to 
ligand induced dewetting. The quantitative discrepancy is probably due to a) the approximations in the VISM functional leading to over-stabilization of dry state relative to wet state, and b) including only local $G[v]$ minima in the ensemble average while omitting intermediate states of not much higher free energy. We expect the qualitative agreement to improve upon inclusion of interface thermal fluctuations into VISM.
\maketitle